\documentclass[aps,prl,twocolumn]{revtex4-1}

\usepackage{bm}
\usepackage{amsmath}
\usepackage{amssymb}
 \usepackage{latexsym}
\usepackage[pdftex]{graphicx}
\usepackage{float}
\usepackage{caption}

\newcommand{\bk}{\mathbf{k}}

\newcommand{\br}{\mathbf{r}}

\newcommand{\bN}{\mathbf{N}}

\newcommand{\bQ}{\mathbf{Q}}

\begin{document}

\title{Possible high temperature odd parity superconductivity in LaNi$_{7/8}$Co$_{1/8}$O$_3$ (111) bilayers}

\author{Bing Ye, Andrej Mesaros, Ying Ran}
\affiliation{Department of Physics, Boston College, Chestnut Hill, MA 02467, USA}
\date{\today}

\begin{abstract}
Using the functional renormalization group technique we demonstrate a route for potentially high-temperature odd-parity superconductivity in ferromagnetic materials caused by repulsive electron interactions, where the superconducting pairing is driven by charge-density wave fluctuations. Our model is directly applicable to a lightly cobalt-doped $LaNiO_3$ bilayer grown in the $(111)$ direction. As the on-site repulsive interaction grows, a charge-density wave state with a charge pattern that respects all point-group symmetries of the bilayer is replaced by a superconducting state with an f-wave pairing.
\end{abstract}

\pacs{}
\maketitle

Beside the well-known electron superconductivity caused by phonon-mediated attractive interactions\cite{Cooper:2011p7868}, Kohn and Luttinger showed\cite{Kohn:1965p7867} that weak repulsive interactions can lead to superconductivity in Fermi liquids, albeit with low transition temperature. The discovery of high temperature superconductivity with d-wave pairing symmetry in copper-oxide compounds\cite{Bednorz:1988p7869,Tsuei:2000p7870} caused a revolution in quantum condensed matter physics and is a source of heated debate to date. However, it is broadly accepted that the strong repulsive Coulomb interaction-caused antiferromagnetic spin correlations play an important role for the superconducting pairing of electrons\cite{Scalapino:2012p7871}. More recently, new excitement was caused by discovery of high temperature superconductivity in iron based pnictide compounds\cite{Stewart:2011p7872,Scalapino:2012p7871}. The parent compounds there are antiferromagnetically ordered too, but due to only moderate electron interactions are also metallic. 

It has been proposed that the antiferromagnetic fluctuations of the iron pnictides drive the superconducting pairing, which has been demonstrated using various versions of renormalization group techiques\cite{Shankar:1994p7862,Polchinski:1984p7873,Metzner:2012p7861,Platt:2013p7896} in the intermediate coupling regime\cite{3Wang:2009p5591,Chubukov:2009p7882,Chubukov:2008p7878,4Platt:2009p7866,Mazin:2008p7881}. In the past, similar theoretical methods were applied to the Hubbard model on the square lattice, which is relevant for the cuprate materials, and produced the correct $d$-wave pairing symmetry\cite{Honerkamp:2001p5910,Zanchi:2000p7874}. From this theoretical point of view, the antiferromagnetic instability at the bare level eventually contributes to the superconductivity scattering channels at lower energies during the renormalization steps. Comparing with the conventional Kohn-Luttinger superconductivity which occurs only at very low temperatures, this mechanism based on spin-density-wave(SDW) fluctuations could cause high temperature superconductivity.

Odd parity superconductors are interesting quantum states of matter which seized a lot of attention recently, partially due to potential applications in realizing Majorana fermions\cite{Read:2000p5872,Ivanov:2001p5868}. However, the known odd-parity superconductors, such as Sr$_2$RuO$_4$\cite{Maeno:2012p7887,MacKenzie:2003p7885,Nelson:2004p7886}, all have rather low transition temperatures, which presents experimental challenges to investigation of their fundamental properties, as well as to using them in various applications.

In this paper we point out that, analogous to the high temperature SDW-driven spin-singlet superconductivity, \textit{charge-density wave} (CDW) fluctuations in a ferromagnetic metallic system can drive high temperature odd parity superconductivity. In addition, this mechanism is directly applicable to certain transition metal oxide heterostructures, which attracted much theoretical and experimental interest in the past decade\cite{Hwang:2012p7889,Okamoto:2013p7888,1Xiao:2011p7761} as promising hosts for various strongly correlated electronic states. In particular, we expect that the CDW fluctutation driven odd parity superconductivity could occur in the LaNi$_{1-x}$Co$_x$O$_3$ bilayer grown in the $(111)$ direction, with light cobalt doping of $x=1/8$. Transition metal oxide heterstructures grown along the (111) direction have been proposed to host various topological phases\cite{1Xiao:2011p7761,2Yang:2011p6110,Ruegg:2012p7890,Ruegg:2011p6117}. Recently successful experimental synthesis of LaNiO$_3$ (111) bilayers was reported\cite{Middey:2012p7893}. A controllable doping by Co-atoms is expected to be achievable using the currently available experimental techniques.

As shown below, due to a partially filled near-flat band, the electronic structure of LaNi$_{1-x}$Co$_x$O$_3$ at $x=1/8$ is expected to host a fully polarized hexagonal Fermi surface with nesting wavevectors. After ferromagnetism is developed, the leading remaining interaction is the on-site inter-orbital repulsion. Nested Fermi surface is well-known to have tendency towards CDW. Using the functional renormalization group (FRG) method, we demonstrate that the CDW fluctuations drive the SC instability, so that above a critical value of on-site inter-orbital repulsion, the CDW state gives way to a superconducting state which has odd symmetry (e.g., f-wave) since it pairs spin-polarized electrons. Here the FRG method is well suited for investigating electronic systems with intermediate repulsion strengths without any bias towards a particular instability.  
\begin{figure}[ht]
\begin{center}
\includegraphics[width=0.5\textwidth]{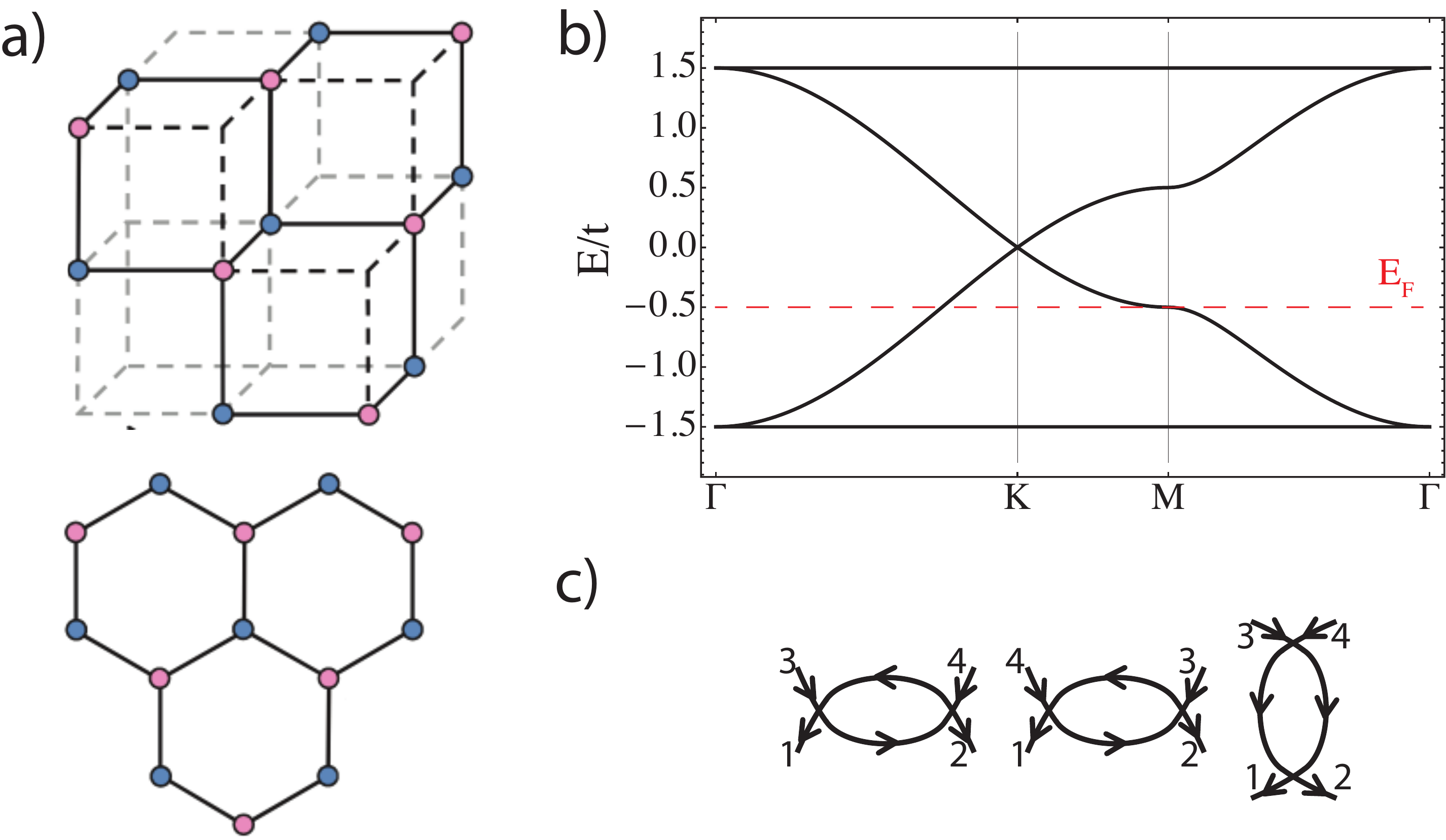}
\end{center}
\caption{(a) The $LaNiO_3$ $(1,1,1)$ bilayer is described by a honeycomb lattice model. (b) The $Co$-doped system is fully spin polarized, having a band structure with two $e_g$ orbitals and two sublattices. (c) Three Feynman diagrams that renormalize the vertex function in FRG. \label{fig1}}
\end{figure}

\textit{Model. ---} Our starting point is the model of $LaNiO_3$ bilayer grown in the $(111)$ direction, which has been described in detail in Refs.\cite{1Xiao:2011p7761,2Yang:2011p6110}. Fig.~\ref{fig1}a shows that the $Ni$ atoms in the two layers form a honeycomb lattice so we use the two-dimensional description of the $[111]$ layer, having a $2\pi/3$ rotational symmetry and a mirror symmetry perpendicular to the plane. In the nearly cubic environment the relevant orbitals are the $e_g$ doublet, to which each nickel ion contributes a single electron, and the resulting nearest neighbor tight-binding band structure in Fig.~\ref{fig1}b has the lower flat bands filled (there are two sublattices, two orbitals and spin degeneracy). After considering further neighbor hoppings the lower flat bands obtain dispersion but is still nearly flat, as shown in a first principle calculation\cite{2Yang:2011p6110}.

\begin{figure}[ht]
\begin{center}
 \includegraphics[width=0.5\textwidth]{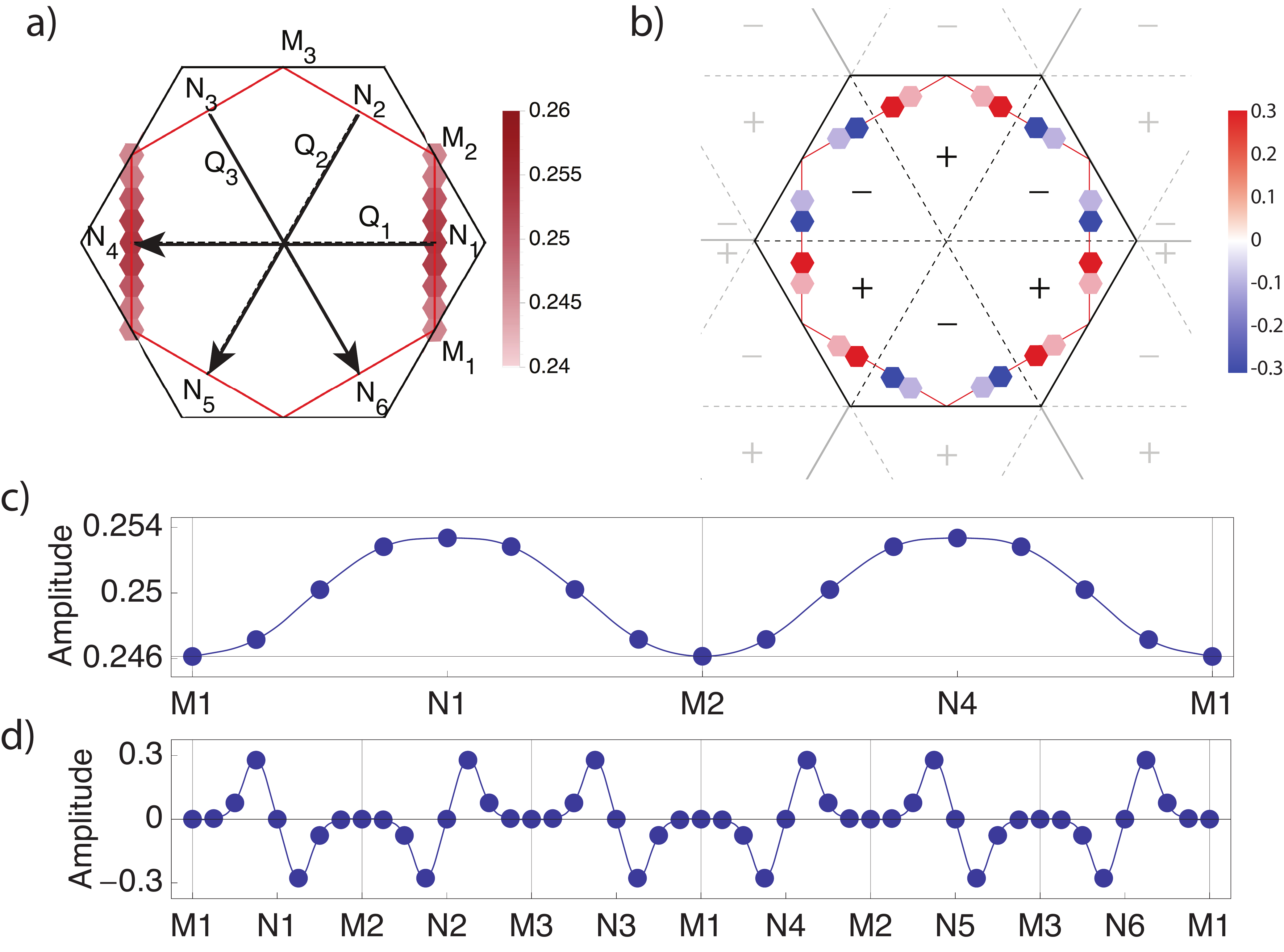}
\end{center}
\caption{(a) The leading charge-density wave instability obtained by FRG. Red line is the Fermi surface, black the Brillouin zone with high symmetry points marked, and the CDW instability form factor is shown for one of the three symmetry related nesting vectors $\bQ$. (c) The form factor plotted as a function along the Fermi surface line. (b) The leading superconducting instability has f-wave pairing. The nodal lines are forced by the Fermi surface nesting. (d) The SC form factor as a function on the Fermi surface line. \label{fig2}}
\end{figure}

In the undoped LaNiO$_3$ bilayer, if interactions are turned off, the Fermi level lies in the middle of a large peak of density of states due to the nearly flat bands. And fully polarized ferromagnetism develops in a Hatree-Fock mean-field calculation\cite{2Yang:2011p6110} even in the presence of moderate repulsions, which can be understood in as a consequence of the Stoner's instability. The cobalt doping is a new ingredient that partially depletes the flat bands, which further enhances the tendency towards ferromagnetism because, when interactions are tuned off, the density of states at the Fermi level is even larger. After turning on the realistic repulsions of the Ni ion, spin-polarized bands are shown in Fig.~\ref{fig1}b, and the $x=1/8$ doping moves the Fermi level from the Dirac touching points to 
the nested Fermi surface depicted in Fig.~\ref{fig2}a. The pertinent model $H=H_{kin}+H_{int}$,
\begin{align}
  \label{eq:1}
  H_{kin}&=-t\sum_{\langle \br , \br' \rangle} \sum_{a,b}  (T^{ab}_{\br\alpha \: \br'\beta} d_{\br, a,\alpha}^\dagger  d_{\br', b,\beta}+h.c.)\\
  H_{int}&= U \sum_{\br,a}\sum_{\alpha<\beta} n_{\br a\alpha} n_{\br a\beta},
\end{align}
includes nearest-neighbor hopping with dimensionless matrix $T$ explicated in Refs.\cite{1Xiao:2011p7761,2Yang:2011p6110}, and an on-site remaining Hubbard repulsion $U$ between the orbitals, with $d_{\br, a,\alpha}$ annihilating a (spin-polarized) electron on honeycomb lattice in unit-cell $\br$, sublattice $a$ and in orbital $\alpha\in\{d_{x^2-y^2},d_{z^2}\}$, with $n_{\br a\alpha}$ being the corresponding electron density. Note that recently various RG techniques were used to propose interesting quantum phases in quarter doped graphene\cite{Raghu:2010p7897,Nandkishore:2012p7877,Wang:2012p7894,Kiesel:2012p7895,Platt:2013p7896}, which has the same Fermi surface as our system (Fig.~\ref{fig2}a), but has spinful electrons and very different orbital structure.

We treat the interaction $H_{int}$ using the FRG method, which provides the renormalized 4-point vertex function $V^\Lambda(\bk_4,b_4;\bk_3,b_3;\bk_2,b_2;\bk_1,b_1)$ as the energy cutoff $\Lambda$ around the Fermi energy is successively shrunk, and less and less states around the Fermi surface are retained. The initial $V^{\Lambda_0}$ is equal to the bare interaction $H_{int}$ expressed in terms of band electron operators $c_{\bk,b}$, with $b$ the band index, where momentum conservation enforces $\bk_3+\bk_4=\bk_1+\bk_2$ and by convention $\bk_1$ and $\bk_4$ belong to operators acting on the same orbital\footnote{The bare interaction, and therefore the initial FRG Hamiltonian, are precisely equal to Eq. (\ref{eq:1}) only if the initial cutoff $\Lambda_0$ is large enough to include the flat bands. However, then there must be a single step in the shrinking of energy cutoff $\Lambda$, no matter how small, which pushes an entire flat band of states outside the cutoff at once. To technically avoid such an FRG step, we set $\Lambda_0<t$ just small enough that the flat bands are completely ignored in the bare interaction $V^{\Lambda_0}$. We then check that FRG results, in particular the CDW and f-wave SC phases, remain robust when the flow is repeated using $V_{corrected}^{\Lambda_0}$, which is also a two-band bare interaction, but which includes a correction obtained by integrating out the flat bands using second-order perturbation theory. This confirms that ignoring the flat bands does not influence the CDW driven SC physics we discuss in this paper.}. In each FRG iteration step the vertex function correction is obtained by summing the Feynman diagrams in Fig.1d. A clear presentation of technical details of FRG is available, e.g., in 
Ref.\cite{Metzner:2012p7861}.
\begin{figure*}[ht]
\begin{center}
   \includegraphics[width=0.9\textwidth]{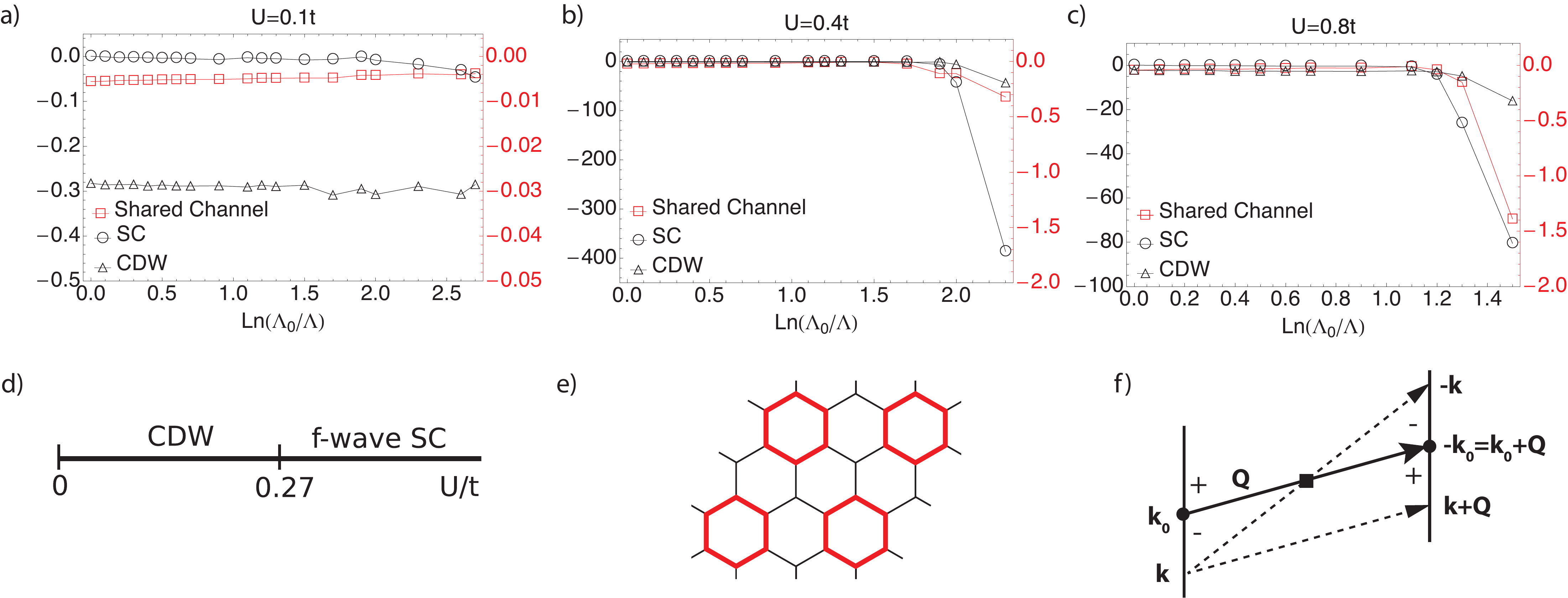}
\end{center}
\caption{(a)-(c) FRG flow of leading CDW and SC instabilities. The shared channel confirms that the initially strong CDW instability drives the SC instability leading to a SC state. (d) The mean field phase diagram obtained from the FRG effective interactions. (e) Schematic of real space CDW pattern which is the mean field ground state in the CDW phase. Sites on red bonds have higher charge density. (f) SC pairing changes sign around special points $\bk_0$ for which the Fermi surface nesting vector $\bQ$ maps $\bk_0$ to $-\bk_0$ (see text).\label{fig3}}
\end{figure*}

At each FRG iteration step, we extract from $V^\Lambda$ the effective interactions that cause CDW and SC instabilities, $V^\Lambda_{CDW}$ and $V^\Lambda_{SC}$, by setting $\bk_4=\bk_1+\bQ$, $\bk_2=\bk_3+\bQ$ and $\bk_4=-\bk_3$, $\bk_1=-\bk_2$, respectively. Due to symmetry, it is enough to consider one CDW nesting vector $\bQ\equiv \bQ_1$ from Fig.~\ref{fig2}a. We consider $16\textrm{x}16$-unit-cell finite periodic system which discretizes lattice momenta $\bk$, and an effective interaction $V^\Lambda(\bk,b;\bk', b')$ becomes a Hermitian matrix with composite indices $(\bk,b)$. The eigenvalues $\lambda_i(\Lambda)$ of this matrix are effective interaction strengths in instability channels $i$, while eigenvectors $v^\Lambda_i(\bk,b)$ are the corresponding form factors\footnote{For simplicity of presentation, this statement does not include the effect of the momentum dependent density of state, which will be taken care of later in the mean-field study.}. For example, at some fixed value of $\Lambda$ (so this index is dropped) the effective interaction for the SC instability becomes $\sum_{i}\lambda_i\hat{\Delta}^\dagger_{SC,i} \hat{\Delta}_{SC,i}$, where $\hat{\Delta}_{SC,
i}\equiv\sum_{\bk,b} v_i(\bk,b) c_{\bk,b} c_{-\bk,b}\equiv\sum_{\bk} \Delta_{SC,i}(\bk)$.

\textit{Results. ---} Figs.~\ref{fig3}a-c show the FRG flow of the most dominant CDW and SC channels with the running cutoff $\ln(\Lambda_0/\Lambda)$, obtained for a range of different interaction strengths $U/t$. The plotted CDW eigenvalue is the most negative and grows the most in the flow, indicating an instability in this CDW channel. The form factor is almost constant and shown in Figs.~\ref{fig2}a,c. \footnote{At the end of the flow in our system with $16\textrm{x}16$ unit-cells, only momenta positioned on the Fermi surface are retained, so the form-factors become functions on the linear segments which define the Fermi surface.} Note that for the smallest interaction strength in Fig.~\ref{fig3}a the divergence of the CDW channel is not developed because the number of fRG steps is limited by finite system size.

For moderate but larger $U/t$ , an SC channel becomes leading during the flow, and its form factor is shown in Figs.~\ref{fig2}b,d --- it is a pairing with f-wave symmetry. Since in the model Eq.~\eqref{eq:1} the electrons are spinless, the pairing must be odd, i.e. $v(\bk,b)=-v(-\bk,b)$, and the pairing form factor has zeros at momenta for which $\bk\equiv -\bk$, the $M$-points on the Fermi surface in Fig.~\ref{fig2}b. However the figure clearly shows that the f-wave pairing in this system is due to a different set of nodal lines where the form factor changes sign. These nodes are caused by the nesting of the Fermi surface, and we will discuss them in detail at the end of the paper.

To confirm the relevance of these instabilities, we combine the kinetic energy $H_{kin}$ from Eq.~(\ref{eq:1}) with the obtained effective interactions and calculate the mean field ground state. Firstly we emphasize that the initial model with the bare interaction has no mean field SC instability for the considered interaction strengths $U/t$. This is consistent with the fact that the lowest SC eigenvalue at the initial step of the flow at any $U/t$ is simply zero. On the other hand, the bare Hamiltonian has a strong instability to CDW due to nesting. The FRG result for the effective interactions $V^{\Lambda_F}_{SC}$ and $V^{\Lambda_F}_{CDW}$, obtained at the end of flow, $\Lambda={\Lambda_F}$, lead to the effective Hamiltonian
\begin{equation}
  \label{eq:2}
  H_{MF}=H_{kin,MF}+H^{\Lambda_F}_{int,MF}
\end{equation}
which is constructed in an energy window $\sim 0.1t$ around the Fermi surface: The $H_{kin,MF}(\bk)$ simply equals $H_{kin}(\bk)$, but we use a bigger lattice ($48\textrm{x}48$ unit-cells) to obtain a good sampling of momenta points $\bk$ within the energy window; the $H^{\Lambda_F}_{int,MF}(\bk)$ at some point $\bk$ is a simple extrapolation of the values of effective interaction $H^{\Lambda_F}_{int}$ which the FRG provides on a smaller set of momenta on the Fermi surface. After the usual mean field treatment of $H_{MF} $, we obtain the phase diagram in Fig.~\ref{fig3}d, which confirms that the CDW ground state is replaced by a SC ground state at $U/t>0.27$. The mean field ground state in the entire CDW phase is described by an equal superposition, with coefficients 1, of the $\bQ_1$, $\bQ_2$ and $\bQ_3$ versions of the CDW form-factor in Fig.~\ref{fig2}a. Fig.~\ref{fig3}e shows a schematic of this ground state in the real space honeycomb lattice, emphasizing the fact that the rotation and mirror symmetries of the lattice are preserved, although the charge pattern doubles the unit-cell in both directions. In the SC phase, the ground state is the f-wave superconductor with the form-factor in Fig.~\ref{fig2}b.

To further reveal how the SC instability prevails in the flow at moderate interaction strengths, in Figure~\ref{fig3} we also present the part of the interaction which is shared by both CDW and SC effective interactions, which is named the ``shared channel''. Namely, if the additional constraint $\bk_3\equiv -(\bk_1+Q)$ is satisfied, the expressions for effective interactions $V_{SC}$ and $V_{CDW}$ become identical and equal to
\begin{equation}
  \label{eq:4}
\hat{V}_{Shared}=\sum_\bk V_{Shared}(\bk,b) c^\dagger_{\bk+\bQ,b}c^\dagger_{-(\bk+\bQ),b}c_{-\bk,b}c_{\bk,b}.  
\end{equation}
The maximal magnitude of the function $V_{Shared}(\bk,b)$ is an indicator of the cooperation between the CDW and SC instabilities, and Figs.~\ref{fig3}a-c show that significant growth of the SC instability correlates with the growth of the shared channel. This confirms our statement that the CDW fluctuations, caused by Fermi surface nesting, drive the SC correlations, allowing for a superconducting state at intermediate Hubbard repulsion strengths.

\textit{Discussion. ---} The SC form factor in Fig.~\ref{fig2}b has an f-wave profile because it changes sign upon crossing nodal lines which pass through high symmetry $\bN$ points on the Fermi surface. We will argue that such SC nodal lines are a generic feature in presence of Fermi surface nesting and inversion symmetry, which provide a simple rule of pairing symmetry of the CDW-driven spin-polarized superconductivity. 

Consider two nested pieces of the Fermi surface which are displaced by a vector $\bQ$, schematically shown in Fig.~\ref{fig3}f. In presence of inversion symmetry, there must be a point $\bk_0$ whose nesting partner $\bk_0+\bQ$ equals its inversion partner $-\bk_0$. We will argue that there is a node in the SC pairing at that point, and furthermore the pairing changes sign on the Fermi surface as $\bk_0$ is crossed (consequently also when $-\bk_0$ is crossed), see Fig.~\ref{fig3}f. We start from the shared channel expression in Eq.~(\ref{eq:4}), which can be written in two ways
\begin{align}
  \label{eq:3}
  \hat{V}_{Shared}&=U\sum_\bk \Delta_{SC}^\dagger(\bk+\bQ) \Delta_{SC}(-\bk)=\\\notag
 &= -U\sum_\bk\Delta_{CDW}^\dagger(-(\bk+\bQ)) \Delta_{CDW}(\bk),
\end{align}
where $\Delta_{CDW}(\bk)=\sum_{b} v^{CDW}(\bk,b) c^\dagger_{\bk+\bQ,b} c_{\bk,b}$ and as earlier $\Delta_{SC}(\bk)=\sum_{b} v^{SC}(\bk,b) c_{\bk,b} c_{-\bk,b}$. The sign in front of $U$ is determined by the rearrangement of fermion operators in the expression for $H_{int}$, and the property $\Delta_{SC}(-\bk)=-\Delta_{SC}(\bk)$ which has been used once.

For the special point $\bk_0$, the interaction term in Eq.~(\ref{eq:3}) becomes $\hat{V}_{Shared} =+U|\Delta_{SC}(-\bk_0)|^2$, and energetically it is obviously preferred for the pairing to vanish at the point $\bk_0$. On the other hand, the same expression equals $\hat{V}_{Shared} =-U|\Delta_{CDW}(\bk_0)|^2$, and the CDW order parameter is not energetically disfavored. Considering a nearby point $\bk$, it is obvious from Fig.~\ref{fig3}f that in the interaction $\hat{V}_{Shared}$ the $\Delta_{SC}$ has a positive coupling to itself (a positive ``mass term'') on the same piece of the Fermi surface but on opposite sides of point $\bk_0$. Energetically therefore the pairing is preffered to change sign at $\bk_0$, as indicated by signs in Fig.~\ref{fig3}f. On the other hand, the CDW order parameter has a negative coupling, and is instead energetically preferred to have the same sign in vicinity of $\bk_0$. Similar analysis was applied previously to study SDW-driven spin-singlet superconductors\cite{Zhai:2009p5911}.

The SC and CDW instabilities we obtained from FRG, Fig.~\ref{fig2}, show this behavior with the six high symmetry points $\bN_a$ taking the role of $\bk_0$. The simple rule established above reproduces the same f-wave pairing symmetry as found in our fRG calculation.

In summary, we discuss a mechanism to realize high temperature odd parity superconducitivity in ferromagnetic metallic systems, driven by CDW fluctutations. This mechanism is directly applicable to the LaNi$_{7/8}$Co$_{1/8}$O$_3$ (111) bilayer oxide heterstructures, in which case $f$-wave superconductivity is found. We also provide a simple rule for the pairing symmetry realized in this mechanism. We hope that this study will encourage experimental growth and characterization of the proposed material and related materials.

We thank Fa Wang for helpful discussions. This work is supported by the Alfred P. Sloan foundation and National Science Foundation
under Grant No. DMR-1151440. We thank Boston College Research Service for providing the computing facilities where the numerical simulations are performed.

\bibliography{fRG_cdwsc}

\end{document}